# Is thermal annealing a viable alternative for crystallization in triethylsilylethynyl anthradithiophene organic transistors?

K. Muhieddine[a,†], R.W. Lyttleton[a,b,†], J. Badcock[a], M.A. Loth[c], J.A. Stride[b], J.E. Anthony[c] and A.P. Micolich[a,*]

[a] School of Physics, The University of New South Wales, Sydney NSW 2052, Australia; [b] School of Chemistry, The University of New South Wales, Sydney NSW 2052, Australia; [c] Department of Chemistry, University of Kentucky, Lexington KY 40506-0055, U.S.A.

[*] Corresponding author. Email: adam.micolich@nanoelectronics.physics.unsw.edu.au

[†] These authors contributed equally to this work

Triethylsilylethynyl anthradithiophene (**TESADT**) holds considerable promise for organic transistor applications due to the high electrical mobilities attained by post-deposition crystallization using solvent vapour annealing. We have studied thermal annealing as an alternative route to post-deposition crystallization of **TESADT** films. Thermal annealing initially appears promising, producing mm-sized crystal domains, but poor electrical performance is obtained, which we attribute to a combination of crack formation and potentially also structural transition during the anneal process. We also find that illumination has a significant positive effect on crystallization, possibly due to an optically-induced enhancement in molecular mobility during annealing. This suggests further studies of how solvent exposure, heat, substrate surface properties and particularly light exposure influence the ordering kinetics of **TESADT** are warranted.

## 1. Introduction

Improved electrical performance is an important goal in the development of organic field-effect transistors (OFETs). Gains are achieved through better molecular ordering in the organic thin-film forming the transistor's conducting channel.[1,2] Carrier mobilities exceeding 0.1 cm$^2$/Vs are routinely obtained using liquid-crystalline semiconducting polymer films, e.g., regioregular polythiophenes,[3] or crystalline small-molecule organics, e.g., pentacene.[4,5] The quest for optimal molecular packing has driven the development of functionalised acenes [6-8] and functionalised acenedithiophenes,[9-12] with triethylsilylethynyl anthradithiophene (**TESADT**) [13] showing particular promise for applications. **TESADT** is readily solution processed into thin films. These films are amorphous upon deposition and easily crystallized by brief exposure to organic solvent vapour, giving saturation electrical mobilities ~0.1 cm$^2$/Vs.[14] **TESADT** films can be patterned [15] using a polydimethylsiloxane (PDMS) stamp prior to annealing or by masked UV illumination during annealing.

Despite **TESADT**'s effectiveness, solvent annealing presents some practical issues for reliable large-area processing; for example, precise areal control over solvent exposure 'dose' is difficult. Thermal annealing is commonly used to improve crystallinity in the fabrication of devices using inorganic semiconductors, e.g., Si.[16,17] Such thermal processes are more readily controlled, and may provide a viable alternative to solvent annealing for organic devices. We investigated this possibility, performing a study of the comparative efficacy of thermal and solvent annealing for improving the order and electrical properties of **TESADT** films. Recent work has also shown that **TESADT** transistors with high electrical mobilities can be obtained by vacuum aging [18] and low temperature casting [19,20] of **TESADT** films. We put our results into context with these novel routes to higher performance **TESADT** transistors, further highlighting the important and complex role that solvent, heat, light and substrate surface play in determining the electronic properties of **TESADT** films.

**2. Materials and Methods**

**TESADT** was synthesized as described previously [9] and a 3% w/w solution in toluene was prepared for spin-coating or drop-casting onto prefabricated, bottom-contact transistor structures. The transistor structures consist of a 10 mm square, heavily-doped Si substrate with 450 nm thermal oxide and six sets of 4 nm Ti : 40 nm Au source/drain contacts deposited using standard photolithographic techniques. The resulting transistor structures have channel width $W = 1000$ μm, channel length $L = 100$ μm, and a specific gate-channel capacitance $C = 7.67 \pm 0.46$ nF/cm$^2$. Substrates were washed in acetone and isopropanol, and O$_2$ plasma cleaned prior to **TESADT** deposition. Spin-coated films were spun at 5,000 rpm for 60 s followed by room temperature vacuum desiccation for 30 min to remove residual solvent. We deposited drop-cast films by touching a 10 μL glass capillary containing **TESADT** solution to the substrate surface. The capillary was withdrawn once solution had spread radially to cover a given transistor structure. The results presented come from study of over a hundred films/devices from five separately synthesised batches of **TESADT**.

Solvent annealing was achieved by a 2 min exposure to dichloroethane vapour.[14] Thermal annealing was performed by placing the substrate on a hotplate at a specific temperature $T$ for a time $t$. Polarized Optical Microscopy (POM) was primarily used to assess crystallinity due to its better ability to demonstrate crystal domain size effects and the large number of samples involved in the study. Shallow angle x-ray diffraction measurements were used to confirm packing structures in cases where this might deviate from known structures in the literature. Electrical measurements were performed in a light-tight, N$_2$ filled, ground-shielded probe station at room temperature. Drain and gate biases $V_d$ and $V_g$ were applied using Keithley 236 source-measure units; these also measure the drain $I_d$ and gate $I_g$ currents. The source current $I_s$ is measured using a Keithley 6517A electrometer, with the source held at ground.

## 3. Results and Discussion

### *3.1 Crystallisation of TESADT thin-films using thermal annealing*

We first determined the upper temperature limit for thermal annealing. The known bulk melting temperature of **TESADT** is 151°C.[13] However, additional structural phase transitions are commonly observed in differential scanning calorimetry (DSC) studies [18,20,21]. Additionally and in particular, a recent study using variable temperature wide-angle x-ray scattering revealed the presence of four polymorphs: one amorphous phase, denoted *a*, and three crystalline phases, denoted α, β and γ.[20] The α-phase is obtained from low temperature casting of the film, the β-phase is obtained at high temperatures ~150°C, and the γ-phase develops from the metastable α-phase after heating and cooling.[20] We will discuss our results in the context of these four phases in more detail below. Figure 1 shows polarized light micrographs of **TESADT** films annealed at 100-120°C for 2 min. The crystal domains decrease in size for $T > 100°C$ with a return to amorphous film structure for $T \geq 115°C$ (Fig. 1(d/e)). Fan-like structures emerge at ~120°C (see Fig. 1(e)); these appear consistent with the high temperature β-phase reported by Yu *et al.*[20] We observe melting of the **TESADT** and dewetting at much higher anneal temperatures. Our films appear black under POM if raised above 136°C. This indicates a total loss of order resulting in amorphous films, consistent with [20].

Figure 2 presents an optimization study with varying time and temperature. **TESADT** films with single crystal domains of area ~1 mm$^2$ are obtained for $T = 80$-100°C, their thermal history and appearance is consistent with the γ-phase.[24] Crystallization is strongly diminished at $T < 80°C$, also consistent with known **TESADT** γ-phase behaviour, i.e., this phase is heat-activated and develops at temperatures above 80°C.[24] For $T > 80°C$, an anneal time $t = 2$ min is sufficient for wide-spread crystallization, $t = 3$-5 min produces more optimum results, and $t > 5$ min produces minimal further improvement. This timescale for completion of crystallization is consistent with solvent anneal studies.[14] The optimum anneal condition was 5 min at 90°C. At its best, our thermal anneal process was able to produce visible crystalline domains at scales matching those obtained from parallel processing using solvent annealing instead, as highlighted in Fig. 3(a/b). This optimum outcome of the thermal anneal tends to vary between **TESADT** batches though, for reasons that we have, as yet, been unable to determine.

Motivated by concerns about photooxidation,[21,22] we also studied how illumination and atmospheric composition affect the outcome of thermal annealing. Interestingly, illumination has a positive effect on crystallization -- films annealed in total darkness show poor crystallization with typical domain size $< 50$ μm whereas films annealed under lab light conditions gave the large domains observed in Figs. 1 and 2. The exact origin is unclear, but may indicate an optically-induced enhancement in molecular mobility during the anneal process. Note that while UV light has previously been used to selectively dewet **TESADT** from $SiO_2$ in the presence of dichloroethane vapour,[15] this occurred with high incident intensity: 540 mW/cm$^2$ at 365 nm for 1-2 min. Dickey *et al.* concluded, based on the intensity used, that dewetting arises from excess thermal energy provided by UV absorption,[15] consistent with the dewetting observed in high temperature anneals, i.e., $T \gg 140°C$. Films thermally annealed under room ambient and pure $N_2$ atmospheres showed little difference in domain size.

These results suggest a complex interplay between solvent interactions and thermal/optical energy during the anneal process. This is also evident in the literature; for example, toluene vapour is an effective annealing agent for **TESADT** [14] and

**TESADT** orders readily thermally, even at room temperature, if given sufficient time.[18] Yet, the solvent removal bake used in earlier work on solvent annealed films, [14,19] which involves both solvent vapour and thermal energy, produces only small-grained polycrystalline films, likely in the γ-phase as this bake is performed at 80-90°C. One possibility is that the thermal and solvent anneals drive towards different structural configurations. To test this we took a solvent annealed film and performed a 2 min thermal anneal at 95°C followed by a 2 min solvent anneal in dichloroethane vapour, and then a second 2 min thermal anneal at 95°C. If the anneals drive towards different configurations then reorganization of the crystal domains would be expected after each anneal. This does not occur. As Fig. 3(c-f) shows, the domain structure is unaffected by subsequent anneals. This behaviour is observed for anneal temperatures up to 120°C where dewetting begins. Shallow angle x-ray diffraction data for solvent and thermal annealed **TESADT** films both show clear, sharp (00ℓ) diffraction peaks, consistent with earlier work.[13,18,23] These results indicate that all films processed at over 80°C and returned to room temperature end up in the γ-phase. Ultimately though, our findings suggest a more detailed study of how solvent exposure, heat, substrate surface properties and particularly light influence the ordering kinetics of **TESADT** is warranted.

*3.2 Electronic properties of thermally annealed TESADT thin film transistors*

Figure 4 shows the drain current $I_d$ versus drain voltage $V_d$ characteristics obtained at a range of gate voltages $V_g$ for thermal and solvent annealed **TESADT** transistors produced using spin-coated (a/b) and drop-cast (c/d) films. The drop-cast **TESADT** transistors were studied in an attempt to lessen the electrical impact of heat-induced film cracking, which we discuss in more detail below. The thicker films obtained by drop-casting still suffer from cracking, but the proportion of the conduction channel cross-section lost to fracture, and the probability that a given fracture will completely sever some segment of the channel, should both be lower. Optimised anneal processes were used in both cases. The films were produced in parallel on nominally identical, pre-prepared transistor substrates: one solvent annealed with dichloroethane vapour for 2 min, the other thermal annealed at 90°C for 5 min. We discuss the spin-coated device results (Fig. 4(a/b)) first. The typical $I_d$ at a given $V_d$ and $V_g$ is at least two orders of magnitude smaller for the thermally annealed device. The electrical characteristics for thermally annealed devices are more erratic and unstable, as is clear for the $V_g = -50$ V (top) trace in Fig. 4(b). For solvent annealed spin-coated films, we obtain a field-effect mobility $\mu \sim 0.1$ cm$^2$/Vs; comparable to previous reports.[14,15,18,24] In contrast, $\mu \sim 2\times10^{-3}$ cm$^2$/Vs is typically obtained for thermally annealed films. Figure 4(c/d) shows the source-drain characteristics for solvent and thermal annealed drop-cast **TESADT** transistors. Both show improved performance compared to the equivalent spin-coated devices. The gains are more significant for the thermally annealed case: the current at a given $V_d$ and $V_g$ is increased by a factor of 2-3, the characteristics look more typical of an FET and show reduced noise. Moving from spin-coated films to drop-cast films approximately doubles the mobility for both anneal types, with $\mu \sim 4\times10^{-3}$ cm$^2$/Vs typical for thermally annealed samples and $\mu \sim 0.2$ cm$^2$/Vs for solvent annealed samples.

    In Fig. 5 we plot the corresponding transfer characteristics ($I_d$ vs $V_g$ at fixed $V_d = -30$ V) for the four devices in Fig. 4. Our devices are considerably less stable under changing $V_g$, making it difficult to obtain high quality transfer characteristics – the resolution of the data in Fig. 5 varies from trace to trace due to persistent issues with device stability. To ensure the obtained transfer characteristics are consistent with the source-drain characteristics, we plot data points obtained at $V_d = -30$ V for the six

different $V_g$ values from Fig. 4 in Fig. 5 also; the correspondence here is strong. The most obvious aspect of Fig. 5 is that the difference between anneal method is much more significant than the difference between deposition method. The solvent annealed films have a much higher on current, consistent with a higher mobility. They also have more positive threshold voltage, a much lower on-off ratio and a significantly reduced subthreshold slope – the values we obtain for these various device performance metrics are compiled in Table 1.

While it is tempting to attribute the poorer performance in our thermally annealed films to reduced crystal domain size and the reduced electrical performance of the TESADT γ-phase compared to other phases, e.g., the α-phase,[20] we believe there is an additional and potentially more significant effect responsible -- thermally induced cracking of the film during the thermal anneal process. We studied this using a polarized optical microscope with a hotplate stage. This enabled us to continuously monitor the thermal anneal process. Visible cracks develop at $T > 75°C$ and persist after cooling to room temperature. To demonstrate the problem most clearly, Figs. 6(a/b) show images of a solvent annealed **TESADT** film immediately before (a) and immediately after (b) thermal annealing at 90°C for 5°min. The post-annealed film in Fig. 6(b) shows numerous cracks, highlighted by the arrows, which were not present in the pre-annealed film. Note well that this problem is not isolated to previously solvent annealed films; amorphous, as-deposited films show the same behaviour, but the cracks are more difficult to see due to the much higher density of grain boundaries in *a*-phase films. There is also no visible change in domain structure between Fig. 6(a) and Fig. 6(b), suggesting that if a change in crystallite phase does occur with solvent annealing, it is invisible to POM. Together these two observations strongly point to thermal expansion being the leading cause of film cracking. However further studies, e.g., by variable temperature wide-angle x-ray scattering, would be needed to comprehensively rule out the possibility that polymorph transitions might accelerate cracking in crystalline films. As a final note, we have been unable to remove these cracks by subsequent solvent or thermal annealing; they appear to be a permanent feature of films annealed to T > 75°C.

We finish by briefly discussing the effect of reduced thermal anneal temperature on the electrical properties of drop-cast **TESADT** transistors to provide some further confirmation that the cracks that develop for thermal anneals at $T > 75°C$ are responsible for the reduced mobility. Figure 6(c) shows the measured mobility vs maximum anneal temperature for a single device subject to anneals at increasing temperatures from 30°C to 90°C. To be specific, we obtain electrical measurements at 25°C, anneal for 2 min at 30°C, obtain electrical measurements at 25°C, anneal for a further 2 min at 40°C, measure again at 25°C, etc., continuing with anneal temperature increments of 10°C until we reach 90°C. We do this in a single device to avoid device-device variations from influencing the results. We find a small increase in mobility with thermal annealing until the anneal temperature exceeds 40°C. The mobility declines thereafter, dropping precipitously for $T > 70°C$, where thermally-induced cracking becomes more prevalent providing confirmation that film cracking adversely affects electrical performance.

The data in Fig. 6(c) suggests that one possible idea to salvage the use of thermal annealing, namely performing much longer anneals (~ hours) at temperatures well below 75°C, is likely of limited potential. The temperature would need to be below 40°C according to Fig. 5(c), and at that point, one might as well either wait the 3-7 days required for spontaneous order to develop at room temperature,[18] resort to solvent annealing,[14] or take what currently appears to be the most favourable option, which is low temperature casting of the **TESADT** film to obtain the metastable α-phase, which gives more consistently favourable electrical characteristics compared to the γ-phase obtained in transistors with **TESADT** films that have undergone high temperature processing.[19,20]

## 4. Conclusions

We studied thermal annealing for post-deposition crystallisation of **TESADT** films for organic transistor applications. Under ideal circumstances thermal annealing produces mm-sized domains comparable in size to established solvent annealing methods.[14] The ideal anneal condition was 5 min at 90°C. The thermal anneal crystalline packing matches that for the solvent anneal, which is likely the TESADT g-phase based on thermal history and appearance under polarized optical microscopy.[20] Repeated annealing by either approach produces no further change in domain structure. Unfortunately, the thermal anneal gives much less improvement in the electrical performance compared to solvent annealing. For spin-coated films we obtained mobilities of order $2\times10^{-3}$ and 0.1 cm$^2$/Vs for thermal and solvent annealed devices, respectively. Both mobility values approximately double for drop-cast films. We attribute the comparatively poor electrical performance of thermally annealed **TESADT** to film cracking during the thermal anneal process. This occurs over and above any polymorph related performance losses.[20] Polarized optical microscopy reveals that cracking occurs above 75°C; these cracks persist upon cooling and even if the film is solvent annealed thereafter. Electrical measurements suggest this problem begins at temperatures as low as 40°C; the cracks developed for 40°C < $T$ < 75°C are likely too small to clearly observe by optically. The cracking is likely due to the difference in thermal expansion coefficient between the **TESADT** and the underlying SiO$_2$/Si substrate. It might be resolved using either flexible substrates (e.g., polyimide films) or substrates with a similar thermal expansion coefficient to **TESADT**.

As a final point, we also note that illumination has a significant positive effect on crystallization with thermal annealing. Films annealed in total darkness show poor crystallization with typical domain size < 50 µm whereas films annealed under lab light conditions give domains up to 1 mm wide. This may indicate an optically-induced enhancement in molecular mobility during annealing, and suggests further studies of how solvent, heat, substrate surface properties, and in particular, light exposure influence the ordering kinetics of **TESADT** are warranted.

**Acknowledgements**

APM acknowledges the Australian Research Council (FT0990285) and JEA acknowledges the National Science Foundation (DMR1035257) for support. This work was performed in part using the NSW node of the Australian National Fabrication Facility (ANFF). We thank A.R. Hamilton and P. Meredith for helpful discussions.

**References**

[1] Sirringhaus H, Brown PJ, Friend RH, Nielsen MM, Bechgaard K, Langeveld-Voss BMW, Spiering AJH, Janssen RAJ, Meijer EW, Herwig P, De Leeuw DM. Two-dimensional charge transport in self-organized, high-mobility conjugated polymers. Nature. 1999;401:685-688.

[2] Gelinck GH, Geuns TCT, de Leeuw DM. High-performance all-polymer integrated circuits. Appl. Phys. Lett. 2000;77:1487-1489.

[3] Baude PF, Ender DA, Haase MA, Kelley TW, Muyres DV, Theiss SD. Pentacene-based radio-frequency identification circuitry. Appl. Phys. Lett. 2003;82:3964-3966.

[4] McCulloch I, Heeney M, Bailey C, Genevicius K, MacDonald I, Shkunov M,

Sparrowe D, Tierney S, Wagner R, Zhang W, Chabinyc ML, Kline RJ, McGehee MD, Toney MF. Liquid-crystalline semiconducting polymers with high charge-carrier mobility. Nat. Mater. 2006;5:328-333.

[5] Anthony JE. The Larger Acenes: Versatile Organic Semiconductors. Angew. Chem. Int. Ed. 2008;47:452-483.

[6] Anthony JE, Brooks JS, Eaton DL, Parkin SR. Functionalized Pentacene: Improved Electronic Properties from Control of Solid-State Order. J. Am. Chem. Soc. 2001;123:9482-9483.

[7] Miao Q, Chi X, Xiao S, Zeis R, Lefenfeld M, Siegrist T, Steigerwald ML, Nuckolls C. Organization of Acenes with a Cruciform Assembly Motif. J. Am. Chem. Soc. 2006;128:1340-1345.

[8] Li Y, Wu Y, Liu P, Prostran Z, Gardner S, Ong BS. Stable Solution-Processed High-Mobility Substituted Pentacene Semiconductors. Chem. Mater. 2007;19:418-423.

[9] Payne MM, Odom SA, Parkin SR, Anthony JE. Stable, Crystalline Acenedithiophenes with up to Seven Linearly Fused Rings. Org. Lett. 2004;6:3325-3328.

[10] Chen M-C, Kim C, Chen S-Y, Chiang Y-J, Chung M-C, Facchetti A, Marks TJ. Functionalized Anthradithiophenes for Organic Field-Effect Transistors. J. Mater. Chem. 2008;18:1029-1036.

[11] Kim C, Huang P-Y, Jhuang J-W, Chen M-C, Ho J-C, Hu T-S, Yan J-Y, Chen L-H, Lee G-H, Facchetti A, Marks TJ. Novel Soluble Pentacene and Anthradithiophene Derivatives for Organic Thin-Film Transistors. Org. Electron. 2010;11:1363-1375.

[12] Lehnherr D, Hallani R, McDonald R, Anthony JE, Tykwinski RR. Synthesis and Properties of Isomerically Pure Anthrabisbenzothiophenes. Org. Lett. 2012;14:62-65.

[13] Payne MM, Parkin SR, Anthony JE, Kuo C-C, Jackson TN. Organic field-effect transistors from solution-deposited functionalized acenes with mobilities as high as 1 cm$^2$/Vs. J. Am. Chem. Soc. 2005;127:4986-4987.

[14] Dickey KC, Anthony JE, Loo Y-L. Improving Organic Thin-Film Transistor Performance through Solvent-Vapor Annealing of Solution-Processable Triethylsilylethynyl Anthradithiophene. Adv. Mater. 2006;18:1721-1726.

[15] Dickey KC, Subramanian S, Anthony JE, Han L-H, Chen S, Loo Y-L. Large-area patterning of a solution-processable organic semiconductor to reduce parasitic leakage and off currents in thin-film transistors. Appl. Phys. Lett. 2007;90:244103.

[16] Roorda S, Sinke WC, Poate JM, Jacobson DC, Dierker S, Dennis BS, Eaglesham DJ, Spaepen F, Fuoss P. Structural relaxation and defect annihilation in pure amorphous silicon. Phys. Rev. B 1991;44:3702-3725.

[17] Spinella C, Lombardo S, Priolo F. Crystal grain nucleation in amorphous silicon. J. Appl. Phys. 1998;84:5383-5414.[18] Lee WH, Lim JA, Kim DH, Cho JH, Jang Y, Kim YH, Han JI, Cho K. Room-Temperature Self-Organizing Characteristics of Soluble Acene Field-Effect Transistors. Adv. Funct. Mater. 2008;18:560-565.

[19] Yu L, Li X, Pavlica E, Loth MA, Anthony JE, Bratina G, Kjellander C, Gelinck G, Stingelin N. Single-step solution processing of small-molecule organic semiconductor field-effect transistors at high yield. Appl. Phys. Lett. 2011; 99:263304.

[20] Yu L, Li X, Pavlica E, Koch FPV, Portale G, da Silva I, Loth MA, Anthony JE, Smith P, Bratina G, Kjellander BKC, Bastiaansen CWM, Broer DJ, Gelinck GH, Stingelin N. Influence of solid-state microstructure on the electronic performance of 5,11-Bis(thiethylsilylethynyl) Anthradithiophene. Chem. Mater. 2013; 25:1823-1828.

[21] Chung YS, Shin N, Kang J, Jo Y, Prabhu VM, Satija SK, Kline RJ, DeLongchamp DM, Toney MF, Loth MA, Purushothaman B, Anthony JE. Zone-refinement effect in small molecule-polymer blend semiconductors for organic thin-film transistors. J. Am. Chem. Soc. 2011;133:412-415.

[22] Subramanian S, Park SK, Parkin SR, Podzorov V, Jackson TN, Anthony JE. Chromophore fluorination enhances crystallization and stability of soluble anthradithiophene semiconductors. J. Am. Chem. Soc. 2008;130:2706-2707.

[23] Dickey KC, Smith TJ, Stevenson KJ, Subramanian S, Anthony JE, Loo Y-L. Establishing Efficient Electrical Contact to the Weak Crystals of Triethylsilylethynyl Anthradithiophene. Chem. Mater. 2007;19:5210-5215.

[24] Lee WH, Kim DH, Cho JH, Jang Y, Lim JA, Kwak D, Cho K. Change of molecular ordering in soluble acenes via solvent annealing and its effect on field-effect mobility. Appl. Phys. Lett. 2007;91:092105.

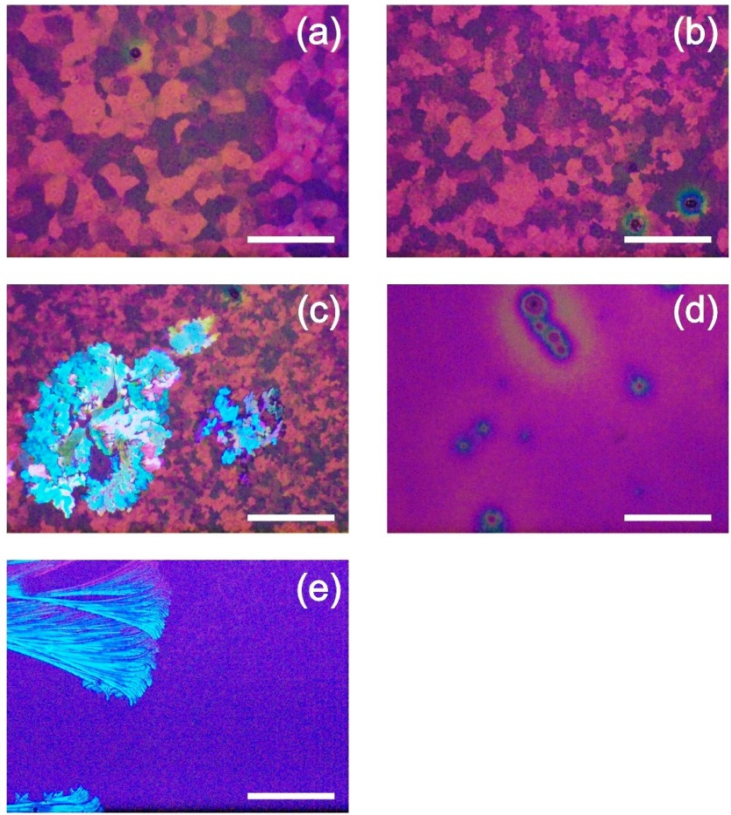

Figure 1. Polarized optical microscopy images for **TESADT** films annealed at $T =$ (a) 100°C, (b) 105°C, (c) 110°C, (d) 115°C, and (e) 120°C for 2 min. The white scale bar in panels (a-d) indicates 100 μm and (e) indicates 1 mm. The post-anneal crystalline domain size decreases for $T > 100$°C with amorphous films obtained for $T > 115$°C.

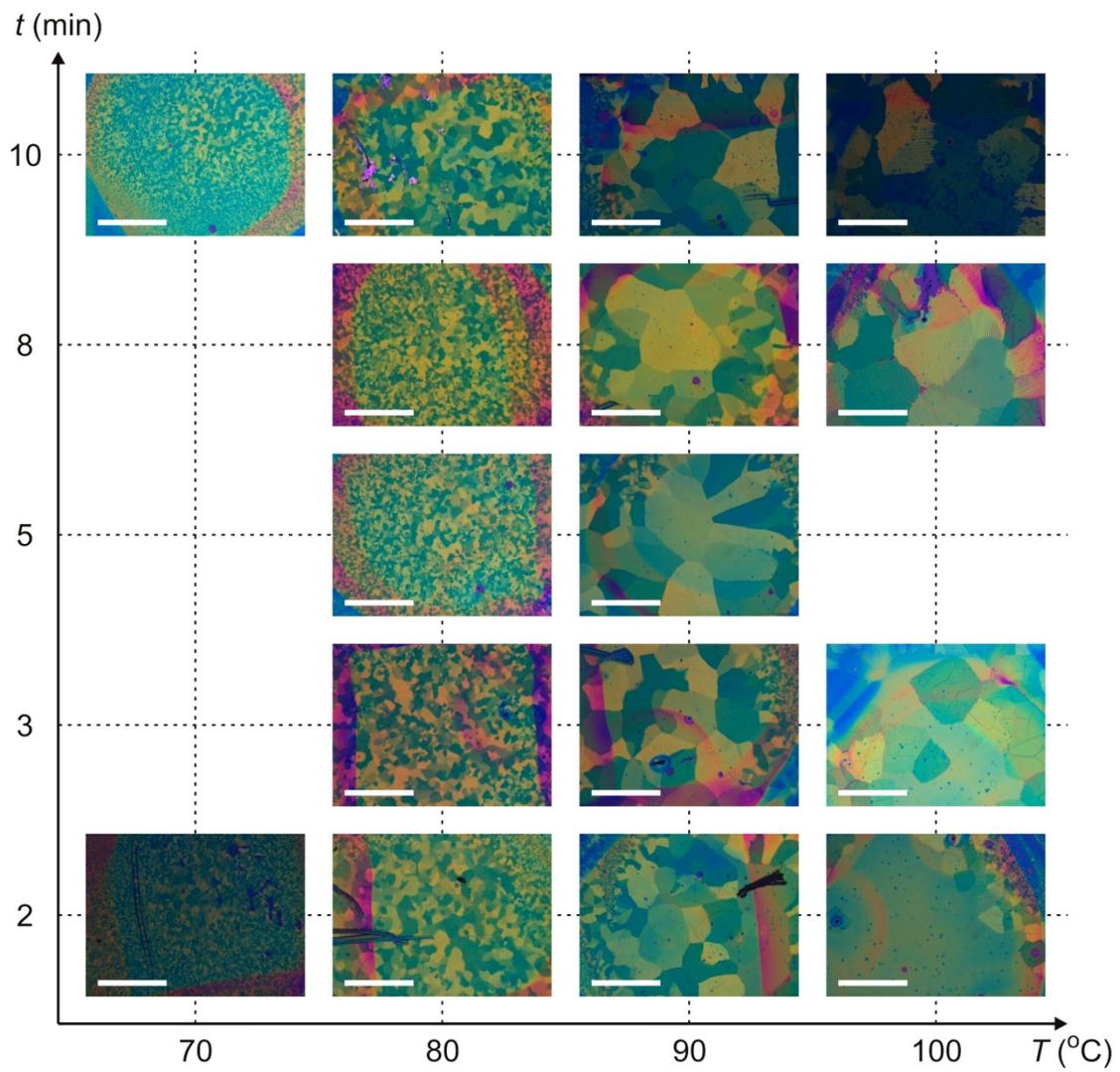

Figure 2. A polarized optical microscopy study of domain size vs thermal anneal temperature $T$ (horizontal axis) and time $t$ (vertical axis) for $T < 100°C$. The white scale bars indicate 1 mm. The optimal anneal conditions are $T \sim 90°C$ for $t \sim 3$ min.

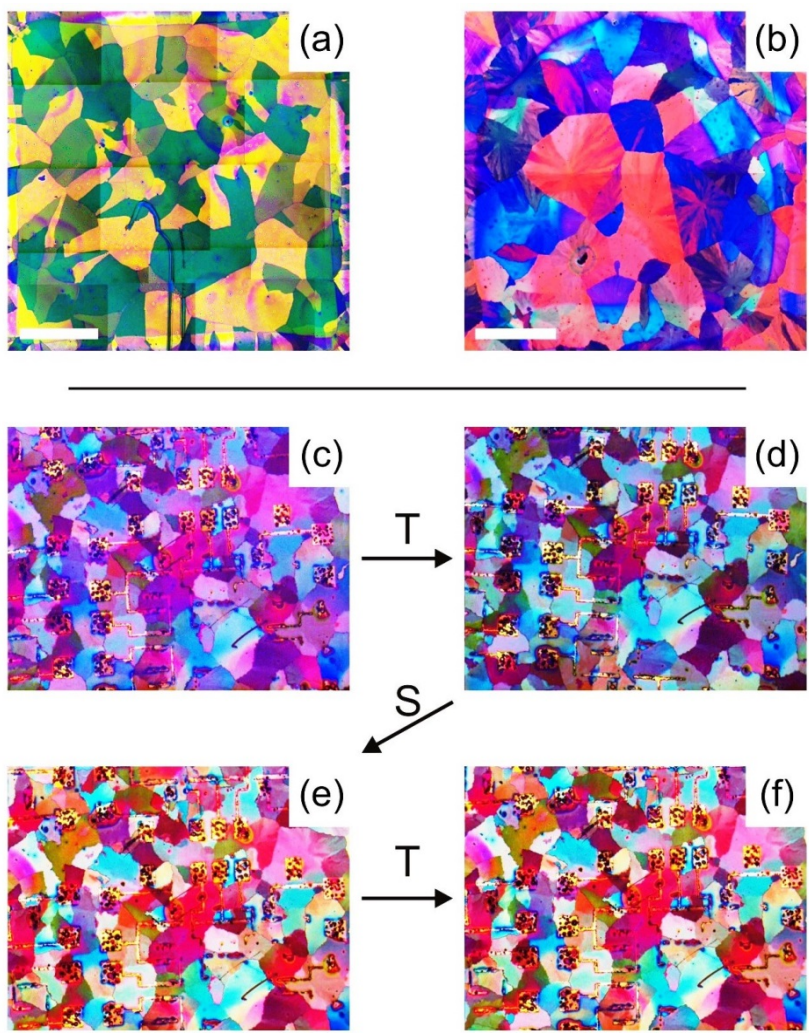

Figure 3. (a,b) Polarized optical microscopy images of the best outcomes obtained using optimised (a) thermal and b) solvent annealing processes. The images are composites of 25 smaller overlapping fields to enable the entire $10 \times 10$ mm area of the chip to be imaged (n.b., the tile-like contrast changes are an artefact). The white scale bars indicate 2 mm. (c-f) Polarized optical microscopy sequence showing the effect of alternating thermal (T) and solvent (S) anneals. The sequence begins with a solvent annealed film (c), the remaining images are obtained after (d) a 2 min thermal anneal at 95°C, (e) a 2 min solvent anneal in dichloroethane vapour, and (f) a second 2 min thermal anneal at 95°C. No visible change in the crystalline domain structure is obtained after the initial anneal confirming that both anneals drive towards the same crystal structure.

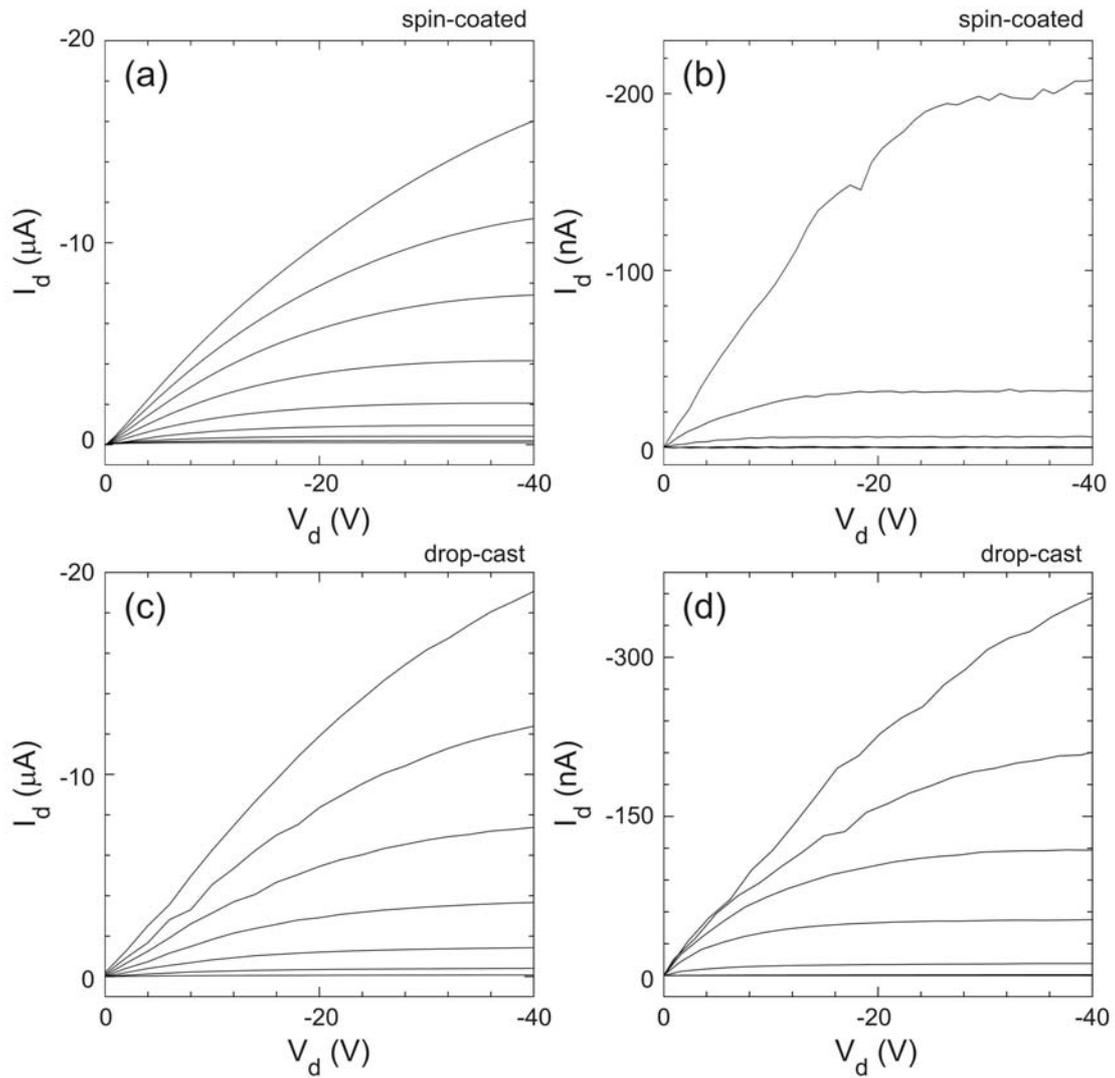

Figure 4. The drain current $I_d$ vs drain voltage $V_d$ for gate voltages from $V_g = -50$ V (top) to 0 V (bottom) in steps of 10 V for (a,b) spin-coated **TESADT** transistors crystallized using (a) a 2 min solvent anneal in dichloroethane vapour, and (b) a 5 min thermal anneal at 90°C, and for (c,d) drop-cast **TESADT** transistors crystallized using (a) a 2 min solvent anneal in dichloroethane vapour, and (b) a 5 min thermal anneal at 90°C.

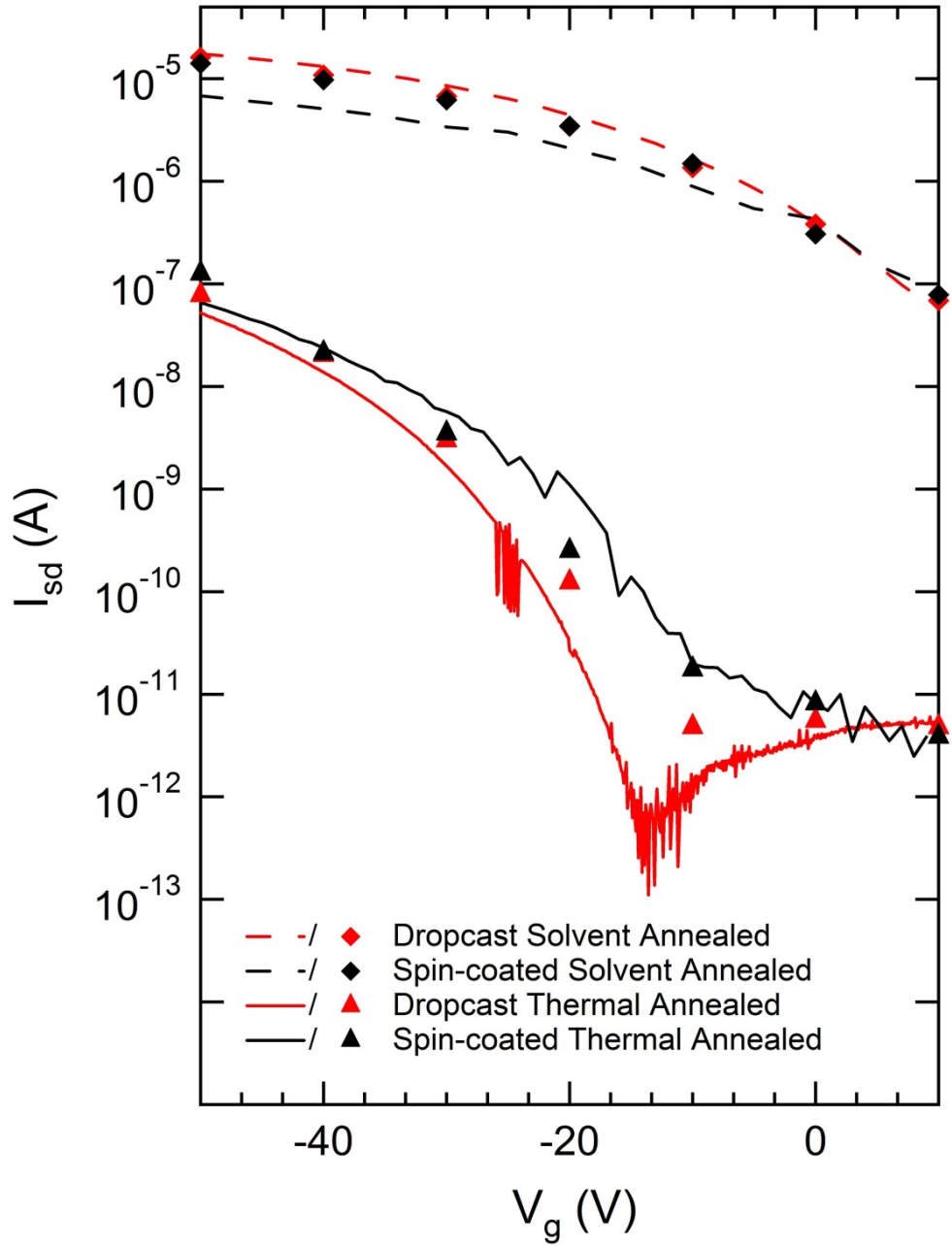

Figure 5. The drain current $I_d$ vs gate voltage $V_g$ for drain voltage $V_d = -30$ V for each of the four different **TESADT** transistors in Fig. 4. The lines are measured transfer characteristics. The data points are extracted directly from Fig. 4 to demonstrate correspondence between the transfer and source-drain characteristics for the four devices.

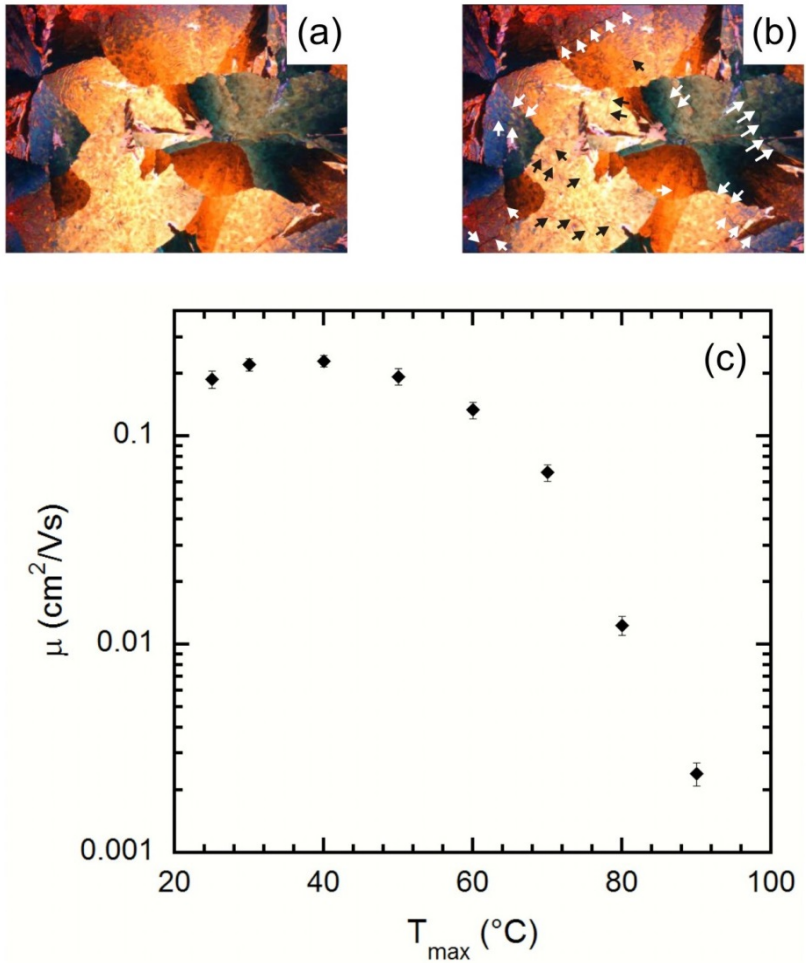

Figure 6. (a,b) Polarized optical microscopy images (a) before and (b) after thermal annealing of a solvent annealed **TESADT** film. The black and white arrows in (b) indicate cracks that developed during thermal annealing. (c) The field-effect mobility μ versus maximum anneal temperature for a **TESADT** transistor that underwent subsequent 2 min thermal anneals at $T = 30, 40, 50, 60, 70, 80$ and $90°C$, with electrical measurements performed at $25°C$ between each anneal. The initial point at $T = 25°C$ indicates a sample measured prior to any thermal annealing.

|  | Device 1 | Device 2 | Device 3 | Device 4 |
| --- | --- | --- | --- | --- |
| Preparation | Spin-coated | Spin-coated | Drop-cast | Drop-cast |
| Anneal | Solvent | Thermal | Solvent | Thermal |
| Mobility (cm$^2$/Vs) | 0.095 | $2.03 \times 10^{-3}$ | 0.18 | $4.13 \times 10^{-3}$ |
| On-off ratio | 82 | 3350 | 284 | 9030 |
| Threshold voltage (V) | −6 | −35 | −11 | −39 |
| Subthreshold swing (V/dec) | 13.8 | 5.6 | 10.6 | 3.5 |

Table 1. Key electrical performance parameters for the four types of TESADT thin-film transistors studied.